\documentstyle[multicol,aps,epsfig]{revtex}

\begin{document}

\draft

\title{Stability of an Ultra-Relativistic Blast Wave in an External
Medium with a Steep Power-Law Density Profile}

\author{Xiaohu Wang, Abraham Loeb\footnote{Guggenheim fellow;
currently on sabbatical leave at The Institute for Advanced Study,
Princeton, NJ 08540}}
\address{Astronomy Department, Harvard
University, 60 Garden Street, Cambridge, MA 02138;
xwang,aloeb@cfa.harvard.edu}
\author{and Eli Waxman}
\address{Department of Condensed Matter Physics, Weizmann Institute,
Rehovot 76100, Israel; waxman@wicc.weizmann.ac.il}

\date{\today}

\maketitle

\begin{abstract}

We examine the stability of self-similar solutions for an accelerating
relativistic blast wave which is generated by a point explosion in an
external medium with a steep radial density profile of a power-law
index $>4.134$. These accelerating solutions apply, for example, to
the breakout of a gamma-ray burst outflow from the boundary of a
massive star, as assumed in the popular collapsar model.  We show that
short wavelength perturbations may grow but only by a modest factor
$\lesssim 10$.

\end{abstract}


\begin{multicols}{2}
\narrowtext

\section{Introduction}

The self-similar solutions of relativistic blast waves are of much
interest because of their recent applications to the study of
Gamma-Ray Bursts (GRBs). A sudden release of a large amount of energy
within a small volume results in a strong explosion that drives a
relativistic shock into the surrounding medium. At late times the
blast wave approaches a self-similar phase whereby its speed and the
distribution of the pressure, density, and velocity of the gas behind
the shock front do not depend on the length and time scales of the
initial explosion, but only on the explosion energy and the properties
of the unshocked external medium. The self-similar solutions
describing this phase have been first studied by Blandford and McKee
\cite{BMK76} (hereafter, BMK).  We list their central results, which
are relevant to this paper, in section \S II.A. Note that in the BMK
solution the total energy released in the explosion $E$ is the only
relevant parameter.

The BMK solution is only valid for $k<4$, where $k$ is the
power-law index of the radial density profile of the external
medium, i.e., $\rho_1 \propto r^{-k}$. When $k \ge 4$, the
similarity variable defined by Blandford and McKee \cite{BMK76} is
no longer appropriate. Even in the range $3\le k<4$, the validity
of the BMK solution is not justified, because the mass contained
behind the shock front diverges if the density profile of the
shocked fluid is described by the BMK solution.

The self-similar solutions for steep density profiles with a power-law
index $k \ge 4$ were derived recently by Best \& Sari
\cite{Best2000}\footnote{The more extreme case of an exponential
density profile has been discussed by Perna \& Vietri
\cite{PV02}.}. The derivation of these solutions is similar to that in
the non-relativistic regime.  The self-similar solutions to a
non-relativistic blast wave were discovered independently by Sedov
\cite{Sedov46} , Von Neumann \cite{VonNeumann47}, and Taylor
\cite{Taylor50}. The so-called "Sedov-Von Neumann-Taylor blast wave"
solutions exist only for $k<5$, but Waxman \& Shvarts \cite{Waxman93}
showed that in the range $3\le k<5$ these solutions fail to describe
the asymptotic flow because the energy diverges; instead they found
second-type self-similar solutions for $3<k<5$ as well as for $k \ge
5$. The new class of non-relativistic, self-similar solutions describe
the flow in a limited spatial region $D(t) \le r \le R(t)$, where
$R(t)$ is the shock radius and $D(t)$ coincides with a $C_+$
characteristic so that the flow inside the region $r< D(t)$ does not
affect the flow in the outer self-similar region. The self-similar
solution has to cross the sonic line into the region where the $C_+$
characteristic can not catch-up with the shock front. The solution
describes a shock accelerating with a temporal dependence whose
power-law index is uniquely determined by requiring that the
self-similar solution cross the sonic line at a singular point.
Note that in these second-type self-similar solutions the total energy
released in the explosion $E$ is not a relevant parameter.  Although
the energy in the self-similar part of the flow approaches a constant
as time diverges, the fraction of the explosion energy $E$ carried by
the self-similar component depends on the details of the initial
conditions. Thus, contrary to the BMK case, dimensional arguments can
not be used to determine the power-law index of the temporal
dependence. Instead, the singular point determines the temporal
power-law index.  In the ultra-relativistic regime, the second-type
self-similar solutions for $k \ge 4$ can similarly be obtained by
requiring that they cross the sonic line at a singular point. Best \&
Sari \cite{Best2000} found that these self-similar solutions exist for
$k>5-\sqrt{3/4}$ and describe accelerating shock waves.  However, the
properties of the flow in the self-similar region, such as the energy
and mass contained in the region, were not discussed.

In this paper, we rederive the self-similar solutions of
ultra-relativistic blast waves for $k>4$ using a different
self-similar variable and discuss the properties of the flow in the
self-similar regime. Our main goal is to study the {\it stability} of
these self-similar solutions.  The stability of the Waxman-Shvarts
self-similar solutions in the non-relativistic regime was studied by
Sari, Waxman \& Shvarts \cite{Sari2000}.  They found that shocks
accelerating at a rate larger than a critical value and corresponding
to solutions that diverge in finite time, are unstable for small and
intermediate wavenumbers.  Shocks that accelerate at a rate smaller
than the critical rate are stable for most wavenumbers. The
acceleration rate can be quantified by the measure $\delta= R \ddot{R}
/ \dot{R}^2$, where the dots denote time derivatives and $R(t)$ is the
radius of the shock front. This measure provides the fractional change
of the velocity over a characteristic time scale for evolution
($R/\dot{R}$). Solutions that diverge in finite time have $\delta>1$
while others have $\delta<1$. Thus, when shocks accelerate
sufficiently fast they become unstable.

In the following sections we study the stability of the self-similar
solutions of ultra-relativistic blast waves for steep density profiles
with a power-law index $k>4$.  The self-similar solutions are
described in \S II. We list the BMK solutions for $k<4$ in \S II.A and
derive the self-similar solutions for $k>4$ in \S II.B.  In \S III, we
discuss the properties of the self-similar flow and calculate the
energy and mass contained in the self-similar regime. In \S IV and \S
V, we study the stability of the self-similar solutions. Finally, we
summarize our main results in \S VI.

\section{Self-Similar Solutions}
\subsection{BMK solutions for $k<4$}

For pedagogical reasons, we first briefly outline the derivation
of the self-similar solutions of relativistic blast waves for
$k<4$ by Blandford \& McKee. For a complete derivation, the reader
is referred to the original paper \cite{BMK76}.

Assuming an ultra-relativistic equation of state, $p=(1/3)e$,
where $p$ and $e$ are the pressure and energy density measured in
the fluid frame, the equations describing a relativistic,
spherically-symmetric, perfect fluid can be written as,
\begin{equation}
\frac{d}{dt}(p\gamma^4)=\gamma^2 \frac{\partial p} {\partial t} ,
\label{eq:fluideqn1}
\end{equation}
\begin{equation}
\frac{d}{dt}\ln (p^3\gamma^4)=-\frac{4}{r^2} \frac{\partial}
{\partial r}(r^2\beta) , \label{eq:fluideqn2}
\end{equation}
\begin{equation}
\frac{\partial n'}{\partial t} + \frac{1}{r^2} \frac{\partial}
{\partial r} (r^2 n' \beta) =0 , \label{eq:fluideqn3}
\end{equation}
where $n'$ is the density as measured in the laboratory frame,
$\gamma$ and $\beta$ are the Lorentz factor and velocity of the
fluid, and
\begin{equation}
\frac{d}{dt} \equiv \frac{\partial} {\partial t} + \beta
\frac{\partial} {\partial r} \label{eq:convderv}
\end{equation}
is the convective derivative. Throughout this paper we set the
speed of light $c$ to unity. Assuming that the blast wave is
ultra-relativistic so that the Lorentz factor of the shock front
$\Gamma$ and the shocked fluid $\gamma$ are much larger than
unity, we only search for solutions accurate to the lowest order
in $\gamma^{-2}$ and $\Gamma^{-2}$.

The effective thickness of the blast wave is approximately
$R/\Gamma^2$, where $R$ is the radius of the shock front. Thus an
appropriate choice of similarity variable is
\begin{equation}
\xi =\left(1- \frac{r}{R} \right)\Gamma^2 \ge 0 . \label{eq:xi}
\end{equation}
Next we assume that the external medium has a scale-free,
power-law density profile $\rho_1 \propto r^{-k}$. Ignoring
radiative losses, the total energy contained in the shocked fluid
remains constant and so the Lorentz factor of the shock front
evolves adiabatically as a power law,
\begin{equation}
\Gamma^2 \propto t^{-m}, \ \ m>-1 . \label{eq:powerlaw}
\end{equation}
Keeping only terms up to order $O(\Gamma^{-2}t)$, the shock radius
is then given by
\begin{equation}
R=t\left[1-\frac{1}{2(m+1)\Gamma^2}\right]. \label{eq:RBMK}
\end{equation}
A more convenient similarity variable can be defined as
\begin{equation}
\chi=1+2(m+1)\xi=[1+2(m+1)\Gamma^2] \left(1- \frac{r}{t} \right) .
\label{eq:defchi}
\end{equation}
In terms of $\chi$, the pressure, velocity, and density in the
shocked fluid can be written as
\begin{equation}
p=\frac{2}{3}w_1\Gamma^2 f(\chi) , \label{eq:p_defBMK}
\end{equation}
\begin{equation}
\gamma^2=\frac{1}{2}\Gamma^2 g(\chi) , \label{eq:gamma_defBMK}
\end{equation}
\begin{equation}
n'=2n_1\Gamma^2 h(\chi) , \label{eq:n_defBMK}
\end{equation}
where $\chi \ge 1$, $w_1$ and $n_1$ are the enthalpy and number
density of the unshocked external medium. We assume that the
unskocked external medium is cold, so that $w_1$ equals the energy
density $\rho_1$. The jump conditions for a strong
ultra-relativistic shock are satisfied by the boundary conditions
\begin{equation}
f(1)=g(1)=h(1)=1 . \label{eq:jumpBMK}
\end{equation}
For an adiabatic impulsive blast wave, equations
(\ref{eq:fluideqn1})--(\ref{eq:fluideqn3}) admit a simple
analytical solution, first derived by BMK \cite{BMK76}
\begin{equation}
f=\chi^{-(17-4k)/(12-3k)} , \label{eq:fBMK}
\end{equation}
\begin{equation}
g=\chi^{-1} , \label{eq:gBMK}
\end{equation}
\begin{equation}
h=\chi^{-(7-2k)/(4-k)} , \label{eq:hBMK}
\end{equation}
for
\begin{equation}
m=3-k>-1 . \label{eq:mkrelBMK}
\end{equation}

\subsection{Self-similar solutions for $k>4$}

In searching for self-similar solutions for $k>4$, we assume that
the Lorentz factor of the shock front still obeys a power law,
$\Gamma^2 \propto t^{-m}$ with $m<-1$.  When $m<-1$, the
similarity variable $\chi$ defined in equation (\ref{eq:defchi})
(and used by Best \& Sari \cite{Best2000}) could be negative. For
convenience we will use $\xi$, defined in equation (\ref{eq:xi}),
instead as our similarity variable. If at an initial time $t_0$,
the shock radius is $R_0$ and the Lorentz factor of the shock
front is $\Gamma_0$, then at a later time $t$, to
$O(\Gamma^{-2}t)$ the shock radius is given by
\begin{equation}
R = R_0+t\left[1-\frac{1}{2(m+1)\Gamma^2}\right] -t_0
\left[1-\frac{1}{2(m+1)\Gamma_0^2}\right]. \label{eq:R}
\end{equation}
We can rewrite this equation as
\begin{equation}
R=t\left[1-\frac{1}{2(m+1)\Gamma^2}\right]+a, \label{eq:Rshort}
\end{equation}
where $a$ is a constant dictated by the initial conditions. This equation
for $R$ with $m<-1$ differs from equation (\ref{eq:RBMK}) by a constant
$a$. However, we can choose the initial time $t_0$ such that $a$ is equal
to zero. This is appropriate because of two reasons.  First, the
self-similar solutions are valid at much later times $t >> t_0$, thus the
effect of the special choice of $t_0$ can be ignored. Second, what matters
in the derivation of the self-similar solutions is the derivative of $R$,
instead of $R$ itself. When $a=0$, the similarity variable becomes
\begin{equation}
\xi = \left(1-\frac{r}{R}\right)\Gamma^2 = \Gamma^2 -\frac{r}{t}
\left[\Gamma^2 +\frac{1}{2(m+1)} \right]. \label{eq:xinew}
\end{equation}
Note that we have ignored higher order terms in $\Gamma^{-2}$ in
the above expression.

Similarly to equations (\ref{eq:p_defBMK})--(\ref{eq:n_defBMK}), we write
the pressure, velocity, and density in the shocked fluid as
\begin{equation}
p=\frac{2}{3}w_1\Gamma^2 f(\xi), \label{eq:p_def}
\end{equation}
\begin{equation}
\gamma^2=\frac{1}{2}\Gamma^2 g(\xi), \label{eq:gamma_def}
\end{equation}
\begin{equation}
n'=2n_1\Gamma^2 h(\xi), \label{eq:n_def}
\end{equation}
where $\xi \ge 0$ and the boundary conditions,
\begin{equation}
f(0)=g(0)=h(0)=1, \label{eq:jump}
\end{equation}
correspond to the jump conditions for a strong ultra-relativistic
shock.

We can now treat $\Gamma^2$ and $\xi$ as two new independent
variables in place of $r$ and $t$, and get
\begin{equation}
t\frac{\partial}{\partial t} = -m \Gamma^2 \frac{\partial}
{\partial \Gamma^2} + \left[ \Gamma^2 -\frac{1}{2}\frac{m}{m+1}
-(m+1)\xi \right] \frac{\partial}{\partial \xi}, \label{eq:ddt}
\end{equation}
\begin{equation}
t\frac{\partial}{\partial r} = -\left[ \Gamma^2 +\frac{1}{2(m+1)}
\right] \frac{\partial}{\partial \xi}, \label{eq:ddr}
\end{equation}
\begin{equation}
t\frac{d}{dt} = -m\Gamma^2 \frac{\partial} {\partial \Gamma^2}
-\left[\frac{1}{2} +(m+1)\xi -\frac{1}{g} \right]
\frac{\partial}{\partial \xi}. \label{eq:dt}
\end{equation}
In deriving the above equations, we have assumed that the blast
wave is ultra-relativistic so that $\Gamma >> 1$ and $\gamma >>
1$. Thus we only keep terms of the lowest contribution order in
$\Gamma^{-2}$ and $\gamma^{-2}$.

Substituting equations (\ref{eq:ddt})--(\ref{eq:dt}) into equations
(\ref{eq:fluideqn1})--(\ref{eq:fluideqn3}), we obtain the following
differential equations for $f$, $g$, and $h$:
\begin{eqnarray}
& & 2(3m+k)g + [g+2(m+1)g\xi+2] \frac{1}{f} \frac{d f}{d \xi}
\nonumber \\ & & + 2[g+2(m+1)g\xi-2] \frac{1}{g} \frac{d g} {d
\xi} = 0, \label{eq:similareqn1}
\end{eqnarray}
\begin{eqnarray}
& & 2(5m+3k-8)g + 3[g+2(m+1)g\xi-2] \frac{1}{f} \frac{d f}{d \xi}
\nonumber \\ & & + 2[g+2(m+1)g\xi+2] \frac{1}{g} \frac{d g}{d \xi}
= 0, \label{eq:similareqn2}
\end{eqnarray}
\begin{eqnarray}
& & 2(m+k-2)g + [g+2(m+1)g\xi-2] \frac{1}{h} \frac{d h}{d \xi}
\nonumber \\ & & + 2\frac{1}{g} \frac{d g}{d \xi} = 0.
\label{eq:similareqn3}
\end{eqnarray}

Using a new variable,
\begin{equation}
y=[1+2(m+1)\xi]g, \label{eq:ydef}
\end{equation}
we rewrite equations (\ref{eq:similareqn1})--(\ref{eq:similareqn3})
as follows
\begin{equation}
\frac{1}{g} \left( \frac{1}{f} \frac{d f}{d \xi} \right) = \frac{
2[4(2m+k-2) -(m+k-4)y]} {y^2-8y+4} , \label{eq:similareqnf}
\end{equation}
\begin{equation}
\frac{1}{g} \left( \frac{1}{g} \frac{d g}{d \xi} \right) = \frac{
2[(7m+3k-4) -(m+2)y]} {y^2-8y+4} . \label{eq:similareqng}
\end{equation}
\begin{eqnarray}
\frac{1}{g} & \left( \frac{1}{h} \frac{dh}{d\xi} \right) =
-2[(y^2-8y+4)(y-2)]^{-1}
\nonumber \\  &
\times [(m+k-2)y^2 -(10m+8k-12)y
\nonumber \\  &
+ (18m+10k-16)]  .
\label{eq:similareqnh}
\end{eqnarray}
One solution to the above equations is obtained for $y=1$ and $m=3-k$, and
corresponds to the BMK solution. From the definition of $y$,
$y=[1+2(m+1)\xi]g$, and the requirement that $g$ be positive it follows
that this solution is only valid for $m>-1$, i.e., $k<4$.

In our search for possible solutions with $k>4$, we start by analyzing
equations (\ref{eq:similareqnf})--(\ref{eq:similareqnh}).  The
right-hand-side of these equations diverges to infinity if
$y^2-8y+4=0$. This corresponds to two singular points,
$y_1=4-2\sqrt{3}=0.536$ and $y_2=4+2\sqrt{3}=7.464$. In addition, equation
(\ref{eq:similareqnh}) has another singular point at $y_3=2$. The solution
to equation (\ref{eq:similareqnf}) can bypass the singular points
$y_1$ and $y_2$ if the numerator on the right-hand-side of the equation
vanishes at $y_1$ or $y_2$. This gives
\begin{equation}
m_{1,2}=\frac{8-4k+(k-4)y_{1,2}}{8-y_{1,2}} . \label{eq:mvalue1}
\end{equation}
It is easy to prove that when $m=m_1$, equations (\ref{eq:similareqng}) and
(\ref{eq:similareqnh}) will also bypass the singular point $y_1$ (the
numerator in the right hand side of each equation is equal to zero at
$y_1$). The same is true for $m=m_2$ and the singular point $y_2$. We will
show below that when $m=m_1$ and $k$ is bigger than a critical value $k_c$,
we have $y \le 1$, thus equations
(\ref{eq:similareqnf})--(\ref{eq:similareqnh}) are able to bypass the
singular point $y_1$ and never reach $y_2$ and $y_3$. The critical value
$k_c$ can be calculated by setting $m_1$ equal to $3-k$, the $m$ value
corresponding to a BMK solution. For $y_1=4-2\sqrt{3}$, we get
\begin{equation}
m_1=12\sqrt{3}-20+(3-2\sqrt{3})k .  \label{eq:m1}
\end{equation}
Thus $m_1=3-k$ gives us
\begin{equation}
k_c=5-\frac{\sqrt{3}}{2} = 4.134 \label{eq:kstart}
\end{equation}
The value of $m_1$ corresponding to $k_c$ is $m_{1c} = -2
+\sqrt{3}/2 = -1.134$. Thus when $k>k_c$, we have $m_1<m_{1c}$.

We now prove that when $m=m_1$ and $k>k_c$, we always have $y \le
1$. Using equation (\ref{eq:similareqng}) and the definition of
$y$ in equation (\ref{eq:ydef}) we obtain
\begin{eqnarray}
\frac{dy}{d\xi} & = & 2(m+1)g(\xi) +\frac{y}{g(\xi)}
\frac{dg}{d\xi} \nonumber \\ & = & -\frac{2g(\xi) [y^2 +
(m-3k+12)y -4(m+1)]} {y^2-8y+4} . \label{eq:dydxi}
\end{eqnarray}
When $m=m_1$, the above equation can be rewritten as
\begin{equation}
\frac{dy}{d\xi} = -\frac{2g(y-b)}{y-y_2}, \label{eq:dydxim1}
\end{equation}
where
\begin{equation}
b=4-10\sqrt{3}+2\sqrt{3}k = 1+ 2\sqrt{3}(k-k_c). \label{eq:b}
\end{equation}
When $k>k_c$, we have $b>1$. We also have $g>0$ and $y_2=4+2\sqrt{3}>1$,
and so the right hand side of equation (\ref{eq:dydxim1}) is negative when
$y \leq 1$. Since the boundary condition is $y(\xi=0)=1$, $y(\xi)$ must be
a monotonically decreasing function of $\xi$ with $y(\xi) \leq 1$. The
asymptotic behavior of $y(\xi)$ can be derived as follows. When $\xi$ is
large, $y$ is negative and $|y|$ is large so that equation
(\ref{eq:dydxim1}) can be approximated as
\begin{equation}
\frac{dy}{d\xi} \simeq -2g \simeq -\frac{1}{m_1+1} \frac{y}{\xi} ,
\label{eq:dydxim1asym}
\end{equation}
where we have used the approximation $y \simeq 2(m_1+1)\xi g$ for
large $\xi$. Equation (\ref{eq:dydxim1asym}) yields
\begin{equation}
-y \propto \xi^{-1/(m_1+1)} . \label{eq:yasym0}
\end{equation}
Note that when $k>k_c$, we have $m_1<m_{1c}<-1$. Thus the exponent
in the above power law is always positive.

It can be proven that when $k \le k_c$, equations
(\ref{eq:similareqnf})--(\ref{eq:similareqnh}) can not bypass all the
singular points with either $m_1$ or $m_2$, and so the BMK solution is the
only possible solution. When $k>k_c$, the equations can not bypass all the
singular points with $m_2$. But this by itself is not sufficient for
justifying that $m_1$ is the only viable choice.  What if the solutions
cut-off at some radius before reaching any singular points? We know that in
order not to run into divergences of the energy or mass of the system, the
solutions must be truncated at some radius $r_+$ (or $\xi_+$ in terms of
the similarity variable), which should coincide with a $C_+$ characteristic
line.  The $C_+$ characteristic guarantees that the flow in the inner
region $r<r_+$ will not influence the flow in the self-similar region $r_+ \le
r \le R$. This $C_+$ characteristic should not overtake the shock front in
finite time, otherwise the self-similar region will eventually
disappear. This argument has been applied in the non-relativistic case
\cite{Waxman93}. We will prove below that in order to get to the regime
where the $C_+$ characteristics can not catch the shock front, the solutions
have to pass the singular point $y_1$, making $m_1$ the only viable choice.

First, let us derive the equation for a $C_+$ characteristic. We use $v$ to
denote the fluid velocity in the laboratory frame. The sound speed in the
fluid frame is $u'_s=1/\sqrt{3}$. Thus the sound speed in the laboratory
frame $u_s$ is given by
\begin{eqnarray}
u_s & = & \frac{u'_s+v}{1+vu'_s} = 1 -\frac {\sqrt{3}-1}
{\sqrt{3}+1} \frac{1} {2\gamma^2} \nonumber \\ & = & 1 -\frac
{\sqrt{3}-1} {\sqrt{3}+1} \frac{1} {\Gamma^2g(\xi)} ,
\label{eq:soundspeed}
\end{eqnarray}
where we only keep the first-order term in $\gamma^{-2}$.
Thus, a $C_+$ characteristic is described by
\begin{equation}
\frac{d r_+}{dt}= 1 -\frac {\sqrt{3}-1} {\sqrt{3}+1} \frac{1}
{\Gamma^2g(\xi_+)} . \label{eq:charac1}
\end{equation}
We can rewrite this equation in terms of the similarity variable
$\xi_+$. Using the definition of $\xi$ in equation (\ref{eq:xi}),
and the relations, $dR/dt = 1-1/(2\Gamma^2)$,
$d\Gamma^2/dt=-m\Gamma^2/t$, we obtain
\begin{equation}
\frac{dr}{dt} = 1-\frac{1}{2\Gamma^2} - \frac{(m+1)} {\Gamma^2}
\xi -\frac{t}{\Gamma^2} \frac{d\xi}{dt}, \label{eq:drdt}
\end{equation}
where we only keep the first-order term in $\Gamma^{-2}$.  Substituting
equation (\ref{eq:drdt}) into equation (\ref{eq:charac1}), we get
\begin{equation}
t\frac{d \xi_+}{dt}= \frac {\sqrt{3}-1} {\sqrt{3}+1} \frac{1}
{g(\xi_+)} -(m+1)\xi_+ -\frac{1}{2}, \label{eq:charac2}
\end{equation}
which describes the evolution of a $C_+$ characteristic. We can
further rewrite this equation as
\begin{equation}
t\frac{d\xi_+}{dt}= -\frac{1}{2g(\xi_+)}(y_+ -y_1),
\label{eq:charac5}
\end{equation}
where
\begin{equation}
y_+ = [1+2(m+1)\xi_+]g(\xi_+) . \label{eq:y+}
\end{equation}
Equation (\ref{eq:charac5}) implies that when $y_+>y_1$, the
right-hand-side of the equation is negative and so $\xi_+$ will decrease
with time and the $C_+$ characteristic will approach the shock front. Only
when $y_+<y_1$, $\xi_+$ will increase with time and the $C_+$
characteristic will not overtake the shock front. Also notice that the
self-similar solution has the boundary condition $y(\xi=0) =1 > y_1$. We
thus proved that in order to get to the regime where $C_+$ characteristics
can not overtake the shock front, the self-similar solution must pass
through the singular point $y_1$, and therefore $m_1$ is the only viable
choice.

We can now attempt to obtain the self-similar solutions for
equations (\ref{eq:similareqnf})--(\ref{eq:similareqnh}). For
$m=m_1$, these equations become
\begin{equation}
\frac{1}{g} \left( \frac{1}{f} \frac{d f}{d \xi} \right) = -\frac{
2(m_1+k-4)} {y-y_2} , \label{eq:dfdxim1}
\end{equation}
\begin{equation}
\frac{1}{g} \left( \frac{1}{g} \frac{d g}{d \xi} \right) =
-\frac{2(m_1+2)} {y-y_2} , \label{eq:dgdxim1}
\end{equation}
\begin{equation}
\frac{1}{g} \left( \frac{1}{h} \frac{dh}{d\xi} \right) =
-\frac{2(m_1+k-2) (y-d)} {(y-y_2) (y-2)} , \label{eq:dhdxim1}
\end{equation}
where
\begin{equation}
d= 4+\sqrt{3} +\frac{ 3+2\sqrt{3}} {k+\sqrt{3}-4}. \label{eq:ddef}
\end{equation}
Treating $y$ as the independent variable instead of $\xi$ and making use of
equation (\ref{eq:dydxim1}), equations
(\ref{eq:dfdxim1})--(\ref{eq:dhdxim1}) can be rewritten as
\begin{equation}
\frac{1}{f} \frac{df}{dy} = \frac{m_1+k-4} {y-b} ,
\label{eq:dfdym1}
\end{equation}
\begin{equation}
\frac{1}{g} \frac{dg}{dy} = \frac{m_1+2} {y-b} , \label{eq:dgdym1}
\end{equation}
\begin{equation}
\frac{1}{h} \frac{dh}{dy} = \frac{(m_1+k-2)(y-d)} {(y-2)(y-b)} .
\label{eq:dhdym1}
\end{equation}
The boundary conditions are $f(y=1)=g(y=1)=h(y=1)=1$. Equations
(\ref{eq:dfdym1}) and (\ref{eq:dgdym1}) have the solutions
\begin{equation}
f=\left( \frac{b-y}{b-1} \right)^{m_1+k-4} , \label{eq:fsolution}
\end{equation}
\begin{equation}
g=\left( \frac{b-y}{b-1} \right)^{m_1+2} . \label{eq:gsolution}
\end{equation}
A special case is obtained at $k=6$ ($m_1=-2$) for which
$f(y)=g(y)=1$.  When $b \neq 2$ ($k \neq (15-\sqrt{3})/3$), equation
(\ref{eq:dhdym1}) has the solution
\begin{equation}
h=\left[ \left( \frac{1}{2-y} \right)^{d-2} \left( \frac{b-y}{b-1}
\right)^{d-b} \right]^{(m_1+k-2)/ (2-b)}.\label{eq:hsolution1}
\end{equation}
When $b = 2$ ($k = (15-\sqrt{3})/3$), equation (\ref{eq:dhdym1})
has the solution
\begin{eqnarray}
h & = & (2-y)^{m_1+k-2} \nonumber \\ & & \times \exp \left[ (m_1+k-2) (d-2)
\left( \frac{1}{2-y} -1 \right) \right] .\label{eq:hsolution2}
\end{eqnarray}

In general, the functions $f(\xi)$, $g(\xi)$ and $h(\xi)$ do not admit
simple analytical forms. Their values can be derived numerically from
equations (\ref{eq:fsolution})--(\ref{eq:hsolution2}). For example,
$g(\xi)$ satisfies the implicit algebraic equation
\begin{equation}
g(\xi)=\left( \frac{b-[1+2(m_1+1)\xi]g(\xi)} {b-1}
\right)^{m_1+2}. \label{eq:gxisolution}
\end{equation}
But generally, we can derive the analytical forms for the asymptotic
behaviors of $f(\xi)$, $g(\xi)$ and $h(\xi)$ in the limit of large
$\xi$. In this limit, $y$ is negative and $|y|$ is large, and so equation
(\ref{eq:gsolution}) yields
\begin{equation}
g \simeq \left( \frac{-y}{b-1} \right)^{m_1+2} \simeq \left[
\frac{-2(m_1+1)\xi g} {b-1} \right]^{m_1+2}. \label{eq:gasym1}
\end{equation}
We can solve $g(\xi)$ from the above equation and get
\begin{equation}
g(\xi) \simeq \alpha_g \xi^{-(m_1+2)/(m_1+1)} , \label{eq:gasym2}
\end{equation}
where
\begin{equation}
\alpha_g = \left[ \frac{-2(m_1+1)}{b-1} \right]^{-(m_1+2)
/(m_1+1)}. \label{eq:alphag}
\end{equation}
Using equation (\ref{eq:gasym2}) we obtain the asymptotic form of
$y(\xi)$ for large $\xi$
\begin{equation}
y(\xi) \simeq -\alpha_y \xi^{-1/(m_1+1)} , \label{eq:yasym}
\end{equation}
where
\begin{equation}
\alpha_y = \left[ \frac{-2(m_1+1)}{(b-1)^{m_1+2}}
\right]^{-1/(m_1+1)}. \label{eq:alphay}
\end{equation}
Using equation (\ref{eq:yasym}), we can derive the asymptotic form
of $f(\xi)$ for large $\xi$ from equation (\ref{eq:fsolution}) and
get
\begin{equation}
f(\xi) \simeq \alpha_f \xi^{-(m_1+k-4)/(m_1+1)} , \label{eq:fasym}
\end{equation}
where
\begin{equation}
\alpha_f = \left[ \frac{-2(m_1+1)}{b-1} \right]^{-(m_1+k-4)
/(m_1+1)}. \label{eq:alphaf}
\end{equation}
Similarly, the asymptotic form of $h(\xi)$ for large $\xi$ can be
derived from equations (\ref{eq:hsolution1}) and
(\ref{eq:hsolution2}). We obtain
\begin{equation}
h(\xi) \simeq \alpha_{h} \xi^{-(m_1+k-2)/(m_1+1)},
\label{eq:hasym}
\end{equation}
where
\begin{equation}
\alpha_h= \left[ \frac{-2(m_1+1)}{1-(b-1)^{(m_1+1)(d-2)/(2-b)}}
\right]^{-(m_1+k-2)/(m_1+1)} \label{eq:alphah1}
\end{equation}
if $b\neq 2$, and
\begin{eqnarray}
\alpha_h & = & \left[ \frac{-2(m_1+1)}{(b-1)^{m_1+2}}
\right]^{-(m_1+k-2)/(m_1+1)} \nonumber \\ & & \times \exp
[-(m_1+k-2) (d-2)] \label{eq:alphah2}
\end{eqnarray}
if $b = 2$.

Equation (\ref{eq:charac5}), which describes the evolution of the
$C_+$ characteristic, can also be rewritten using the variable
$y$. Using equation (\ref{eq:dydxim1}), we obtain
\begin{equation}
t\frac{dy_+}{dt}=\frac{(y_+-y_1)(y_+ -b)} {y_+-y_2}.
\label{eq:dy+dtm1}
\end{equation}
We have proven earlier that when $k>k_c$ one finds $b>1$ and $y \leq 1$. Thus
the sign of the right-hand-side of the above equation is decided by the
term $(y_+-y_1)$. If a $C_+$ characteristic emerges from the region
$y<y_1$, i.e., $y_+(t=t_0) < y_1$, we know that $y_+(t)$ will decrease with
time or equivalently $\xi_+(t)$ will increase with time and the $C_+$
characteristic will not catch-up with the shock front.

Equations (\ref{eq:dfdxim1})--(\ref{eq:dhdxim1}) can also be solved
numerically for different values of $k$. We plot the results for $k=5.5$
($m_1=-1.77$) and $k=6.5$ ($m_1=-2.23$) in Figure 1. When $k_c<k<6$, the
function $f(\xi)$ decreases with increasing $\xi$ while the function
$g(\xi)$ increases with increasing $\xi$. This implies that when moving
inwards away from the shock front, the pressure decreases while the Lorentz
factor increases. For $k>6$ the situation is reversed.  For all $k>k_c$,
the function $h(\xi)$ increases with $\xi$ implying that the density always
increases when moving inwards away from the shock front.

\begin{figure}[!t]
\centerline{\psfig{figure=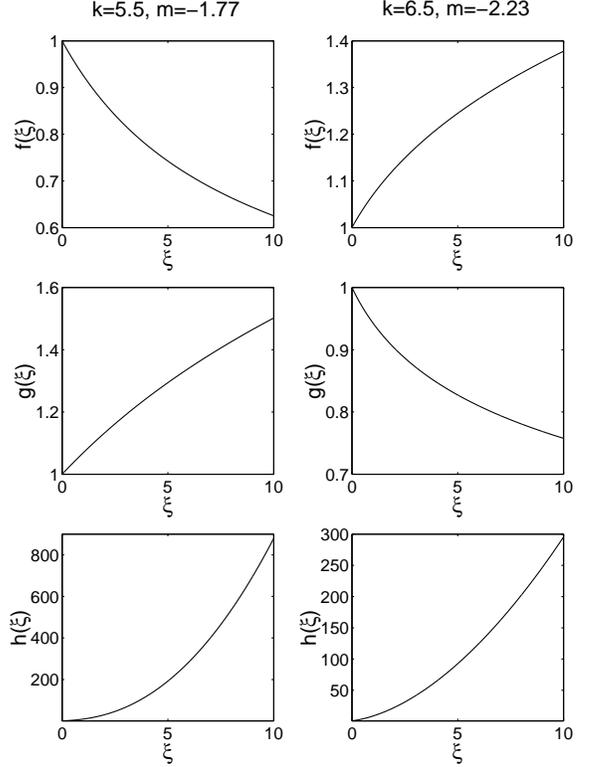,width=3.0in}} \caption[]{
Distributions of the self-similar functions $f(\xi)$, $g(\xi)$ and
$h(\xi)$. The left column corresponds to $k=5.5$ and $m_1=-1.77$, while the
right column corresponds to $k=6.5$ and $m_1=-2.23$.} \label{figure1}
\end{figure}

\section{Properties of the flow in the self-similar region}

We now examine the properties of the flow in the self-similar region
bounded by $\xi=0$ and $\xi=\xi_+(t)$, where $\xi_+(t)$ coincides with a
$C_+$ characteristic that emerges from the region $y<y_1$. The evolution of
the $C_+$ characteristic is described by equation (\ref{eq:dy+dtm1}) with
the initial condition $\xi_+(t=t_0)=\xi_0$, and correspondingly
$y_+(t=t_0)=y_0<y_1$.  By solving equation (\ref{eq:dy+dtm1}), we get the
following equation for $y_+(t)$,
\begin{equation}
\left( \frac{y_1-y_+} {y_1-y_0} \right)^{(y_2-y_1) /(b-y_1)}
\left( \frac{b-y_+} {b-y_0} \right)^{(b-y_2) /(b-y_1)} =
\frac{t}{t_0}. \label{eq:y+analy}
\end{equation}
We can now derive the asymptotic behavior of the $C_+$ characteristic
at $t \rightarrow \infty$. In this limit $|y_+(t)|$ is
large, and equation (\ref{eq:y+analy}) yields
\begin{equation}
y_+(t) \simeq -\alpha_0 t, \label{eq:y+asym}
\end{equation}
where
\begin{equation}
\alpha_0 = (y_1-y_0)^{(y_2-y_1) /(b-y_1)}(b-y_0)^{(b-y_2)
/(b-y_1)} /t_0. \label{eq:alpha0}
\end{equation}
Substituting equation (\ref{eq:yasym}) into equation (\ref{eq:y+asym}), we
get
\begin{equation}
\xi_+(t) \simeq \alpha_+ t^{-(m_1+1)}, \label{eq:xi+asym}
\end{equation}
where
\begin{equation}
\alpha_+ = \left( \frac{\alpha_0} {\alpha_y} \right)^{-(m_1+1)}.
\label{eq:alpha+}
\end{equation}
The initial condition for the $C_+$ characteristic, namely the value of the
$C_+$ characteristic with which $\xi_+(t)$ coincides, contains the
information about the initial explosion.

The shock front is accelerating with a power-law temporal dependence of its
Lorentz factor, $\Gamma^2 \propto t^{-m_1}$. How does the $C_+$
characteristic propagate?  From equations (\ref{eq:gasym2}) and
(\ref{eq:xi+asym}) we find that when $t \rightarrow \infty$,
\begin{equation}
\gamma^2(\xi_+) = \frac{1}{2} \Gamma^2 g(\xi_+) \propto t^2.
\label{eq:c+lorentz}
\end{equation}
We see that irrespective of the value of $k$, the $C_+$ characteristic
always accelerates as $\gamma^2 \propto t^2$. We can also calculate the
thickness of the self-similar region. Using equations (\ref{eq:xi}) and
(\ref{eq:xi+asym}) we obtain when $t \rightarrow \infty$,
\begin{equation}
R-r_+ = \frac{R \xi_+} {\Gamma^2} \longrightarrow {\rm const} .
\label{eq:thickness}
\end{equation}
Note that when $-2<m_1<m_{1c}$, the $C_+$ characteristic accelerates faster
than the shock front, but because the Lorentz factors of both surfaces are
accelerating as power-laws of time, the $C_+$ characteristic can never
catch-up with the shock front. Instead, the distance between the two surfaces
approaches a constant value at late times.

We can now examine the energy and mass contained in the self-similar
region. The energy contained in the spherical shell between $\xi=0$ and
$\xi=\xi_+(t)$ is given by
\begin{equation}
E = \int_{r_+}^{R} 16\pi p\gamma^2r^2dr = \frac{16 \pi}{3} w_1
\Gamma^2 R^3 \int_{0}^{\xi_+} f(\xi) g(\xi)d\xi .
\label{eq:energy}
\end{equation}
Using equations (\ref{eq:fasym}) and (\ref{eq:gasym2}), we can
calculate the above integral for large values of $\xi_+$.
This gives,
\begin{eqnarray}
E & \simeq & \frac{16 \pi}{3} \left( -\frac{m_1+1}{m_1+k-3}
\right) \alpha_f \alpha_g w_1 \Gamma^2 R^3 \nonumber \\
& & \times \xi_+^{-(m_1+k-3) /(m_1+1)}.  \label{eq:fginteg}
\end{eqnarray}
In deriving the above result we have used the fact that when $k>k_c$, we
have $m_1+k-3>0$ and $m_1+1<0$, so that the exponent of $\xi_+$ in the
above equation $-(m_1+k-3)/(m_1+1)$ is positive.  When $t \rightarrow
\infty$, $\xi_+$ is given by equation (\ref{eq:xi+asym}). In addition
$\Gamma^2 \propto t^{-m_1}$, $w_1 = \rho_1 \propto R^{-k}$. Thus when $t
\rightarrow \infty$,
\begin{equation}
E \longrightarrow {\rm const} . \label{eq:Econst}
\end{equation}

The total number of particles contained between $\xi=0$ and $\xi=\xi_+(t)$
is given by
\begin{equation}
N = \int_{r_+}^{R} n' 4\pi r^2 dr = 8\pi n_1 R^3 \int_{0}^{\xi_+}
h(\xi) d\xi . \label{eq:mass}
\end{equation}
Using equations (\ref{eq:hasym}) and (\ref{eq:xi+asym}), we obtain that for
$t \rightarrow \infty$,
\begin{eqnarray}
& N & \simeq 8\pi \left(-\frac{m_1+1} {k-3} \right) \alpha_h n_1
R^3 \xi_+^{-(k-3) /(m_1+1)} \nonumber \\ & & \longrightarrow {\rm
const} . \label{eq:Nconst}
\end{eqnarray}

We have thus proven that both the energy and mass contained between the
$C_+$ characteristic and the shock front will approach constant values as
$t \rightarrow \infty$. The situation is similar to the non-relativistic
case \cite{Waxman93}.

\section{Approximate (Analytic) Stability Analysis}

In order to analyze the stability of the self-similar solutions
obtained in \S II.B, we first follow an analytic approach (in \S
IV) based on the assumptions of variable separation and a fixed
$\Gamma_0$, where $\Gamma_0$ is the unperturbed Lorentz factor of
the shock front. As we will explain later, these assumptions limit
the generality of the results. We then use numerical simulations
(in \S V) to directly solve the evolution of the perturbations without
those assumptions. The numerical simulations demonstrate that the
results obtained using the analytic approach are qualitatively valid.

\subsection{Derivation of linear perturbation equations}

For the analytic approach to the stability analysis of the
self-similar solutions, we use linear perturbation analysis
similar to that used in the non-relativistic case \cite{Sari2000}.
We start from the equations of motion for an ideal relativistic
fluid:
\begin{equation}
\frac{\partial}{\partial t}(n') + {\bf \nabla} \cdot (n' {\bf v})
= 0 , \label{eq:pert_mass}
\end{equation}
\begin{equation}
\gamma^2 (e+p) \left[ \frac{\partial {\bf v}}{\partial t} + ({\bf
v} \cdot {\bf \nabla}){\bf v} \right ] + \left( {\bf \nabla}p +
{\bf v}\frac{\partial p}{\partial t} \right) = 0 ,
\label{eq:pert_momentum}
\end{equation}
\begin{equation}
\frac{\partial}{\partial t} \left( \frac{\gamma^{4/3} p}
{n'^{4/3}}\right) + ({\bf v} \cdot {\bf \nabla}) \left(
\frac{\gamma^{4/3} p} {n'^{4/3}}\right) =0 ,
\label{eq:pert_energy}
\end{equation}
where ${\bf v}$ and $\gamma$ are the fluid velocity and Lorentz
factor in the perturbed solution measured in the laboratory frame,
$e$ and $p$ are the energy density and pressure in the perturbed
solution measured in the fluid frame, and $n'$ is the fluid
density in the perturbed solution measured in the laboratory
frame. We use the Eulerian perturbation approach, i.e., the
perturbed quantities are defined as the difference between the
perturbed solution and the unperturbed one in the same spatial
point. Therefore we define the perturbed hydrodynamic quantities
as
\begin{equation}
\delta p (r, \theta, \phi, t) = p (r, \theta, \phi, t) - p_0(r, t)
, \label{eq:deltap}
\end{equation}
\begin{equation}
\delta {\bf v}(r, \theta, \phi, t) = {\bf v} (r, \theta, \phi, t)
- v_0(r, t) \hat{\bf r} = \delta v_r \hat{\bf r} + \delta {\bf
v}_T , \label{eq:deltav}
\end{equation}
\begin{equation}
\delta n' (r, \theta, \phi, t) = n' (r, \theta, \phi, t) - n'_0(r,
t) , \label{eq:deltan}
\end{equation}
where the quantities with subscript ``0'' are the unperturbed
values. Substituting the above quantities into equations
(\ref{eq:pert_mass})--(\ref{eq:pert_energy}), we obtain the
following linear perturbation equations,
\begin{eqnarray}
& & \frac{\partial}{\partial t} \delta n' + \frac{1}{r^2}
\frac{\partial}{\partial r} \left\{ r^2\left[ \frac{1} {2
\gamma_0^4}\left(1 + \frac{1} {2 \gamma_0^2}\right) n'_o \delta
\gamma^2 \right. \right. \nonumber \\ & & \left. \left. + \left(1
- \frac{1} {2 \gamma_0^2}\right) \delta n' \right] \right\} + n'_0
{\bf \nabla}_T \cdot \delta {\bf v}_T = 0, \label{eq:deltamass}
\end{eqnarray}
\begin{eqnarray}
& & \frac{2}{\gamma_0^4} \left[ \left(1+ \frac{1} {2
\gamma_0^2}\right) \frac{\partial \gamma_0^2}{\partial t} +
\frac{\partial \gamma_0^2}{\partial r} \right] (\gamma_0^2 \delta
p + p_0 \delta \gamma^2 ) \nonumber \\ & & + 4 \gamma_0^2 p_0
\left\{ \frac{\partial}{\partial t} \left[ \frac{1} {2 \gamma_0^4}
\left(1+ \frac{1} {2 \gamma_0^2}\right) \delta \gamma^2 \right]
\right. \nonumber \\ & & + \left(1- \frac{1} {2 \gamma_0^2}\right)
\frac{\partial}{\partial r} \left[ \frac{1} {2\gamma_0^4} \left(1+
\frac{1} {2 \gamma_0^2}\right) \delta \gamma^2 \right] \nonumber
\\ & & + \left. \frac{1} {4 \gamma_0^8} \left(1+ \frac{1} {
\gamma_0^2}\right) \left( \frac{\partial \gamma_0^2}{\partial r}
\right) \delta \gamma^2 \right\} \nonumber
\\ & & + \frac{\partial}{\partial r} \delta p + \left( 1- \frac{1}
{2 \gamma_0^2} \right) \frac{\partial}{\partial t} \delta p
\nonumber \\ & & + \frac{1} {2 \gamma_0^4} \left(1+ \frac{1} {2
\gamma_0^2}\right) \left( \frac{\partial p_0}{\partial t} \right)
\delta \gamma^2 = 0, \label{eq:deltagamma}
\end{eqnarray}
\begin{eqnarray}
& & 4p_0\gamma_0^2 \left[ \frac{\partial}{\partial t} \delta {\bf
v}_T + \left( 1- \frac{1} {2 \gamma_0^2} \right)
\frac{\partial}{\partial r} \delta {\bf v}_T \right. \nonumber \\
& & \left. + \left( 1- \frac{1} {2 \gamma_0^2} \right) \frac{1}{r}
\delta {\bf v}_T \right] + {\bf \nabla}_T \delta p + \left(
\frac{\partial p_0}{\partial t} \right) \delta {\bf v}_T = 0,
\label{eq:deltavt}
\end{eqnarray}
\begin{eqnarray}
& & \frac{\partial}{\partial t} \left[  \frac{\gamma_0^{4/3} p_0}
{n'^{4/3}_0} \left(\frac{2}{3} \frac{\delta \gamma^2} {\gamma_0^2}
+ \frac{ \delta p} {p_0} -\frac{4}{3}\frac{ \delta n'} {n'_0}
\right)\right] \nonumber \\ & & + \left( 1- \frac{1} {2
\gamma_0^2} \right) \frac{\partial}{\partial r} \left[
\frac{\gamma_0^{4/3} p_0} {n'^{4/3}_0} \left(\frac{2}{3}
\frac{\delta \gamma^2} {\gamma_0^2} + \frac{ \delta p} {p_0}
-\frac{4}{3}\frac{ \delta n'} {n'_0} \right)\right] \nonumber \\ &
& + \frac{1}{2\gamma_0^4} \left(1+ \frac{1} {2 \gamma_0^2}\right)
\frac{\partial}{\partial r} \left( \frac{\gamma_0^{4/3} p_0}
{n'^{4/3}_0} \right) \delta \gamma^2 = 0, \label{eq:deltaenergy}
\end{eqnarray}
where we have used the relations
\begin{equation}
\frac{\partial v_0}{\partial t} = \frac{1}{2\gamma_0^4} \left(1+
\frac{1} {2 \gamma_0^2}\right) \frac{\partial \gamma_0^2}{\partial
t} , \label{eq:dv0dt}
\end{equation}
\begin{equation}
\frac{\partial v_0}{\partial r} = \frac{1}{2\gamma_0^4} \left(1+
\frac{1} {2 \gamma_0^2}\right) \frac{\partial \gamma_0^2}{\partial
r} , \label{eq:dv0dr}
\end{equation}
\begin{equation}
\delta v_r = \frac{1}{2\gamma_0^4} \left(1+ \frac{1} {2
\gamma_0^2}\right) \delta \gamma^2 , \label{eq:deltavrgamma}
\end{equation}
and the operator ${\bf \nabla}_T \equiv (\hat{\theta}/r)(\partial
/\partial \theta) + (\hat{\phi}/r)(\partial/\partial \phi)$ acts
as follows on a scalar $\Psi$ and a vector ${\bf f}$:
\begin{equation}
{\bf \nabla}_T\Psi = \frac{1}{r}\frac{\partial \Psi} {\partial
\theta} \hat{\theta} +\frac{1}{r\sin \theta}\frac{\partial \Psi}
{\partial \phi} \hat{\phi} , \label{eq:deltaTgradient}
\end{equation}
\begin{equation}
{\bf \nabla}_T \cdot {\bf f} = \frac{1}{r \sin
\theta}\frac{\partial}{\partial \theta} (\sin \theta f_{\theta})
+\frac{1}{r\sin \theta}\frac{\partial f_{\phi}} {\partial \phi} .
\label{eq:deltaTdivergence}
\end{equation}

Since the unperturbed quantities satisfy equations
(\ref{eq:p_def})--(\ref{eq:n_def}), we write
\begin{equation}
p_0=\frac{2}{3}\rho_1\Gamma_0^2 f(\xi), \label{eq:p_def_unpert}
\end{equation}
\begin{equation}
\gamma_0^2=\frac{1}{2}\Gamma_0^2 g(\xi),
\label{eq:gamma_def_unpert}
\end{equation}
\begin{equation}
n'_0=2n_1\Gamma_0^2 h(\xi), \label{eq:n_def_unpert}
\end{equation}
where $\Gamma_0$ is the unperturbed Lorentz factor of the shock
front, and $\xi$ is the similarity variable defined as
\begin{equation}
\xi =\left(1- \frac{r}{R_0} \right)\Gamma_0^2 ,
\label{eq:xi_unpert}
\end{equation}
where $R_0$ is the unperturbed radius of the shock front.
We further define the perturbation variables as
\begin{equation}
\delta \gamma^2 (r, \theta, \phi, t) = \frac{1}{2} \Gamma_0^2
\delta g (\xi) Y_{lm} (\theta, \phi) X(t), \label{eq:deltagammaY}
\end{equation}
\begin{equation}
\delta {\bf v}_T (r, \theta, \phi, t) = -\frac{1}{\Gamma_0^2}
\delta v_T (\xi) \tilde{ {\bf \nabla}}_T Y_{lm} (\theta, \phi)
X(t), \label{eq:deltavtY}
\end{equation}
\begin{equation}
\delta p (r, \theta, \phi, t) = \frac{2}{3} \rho_1 \Gamma_0^2
\delta f (\xi) Y_{lm} (\theta, \phi) X(t), \label{eq:deltapY}
\end{equation}
\begin{equation}
\delta n' (r, \theta, \phi, t) = 2 n_1 \Gamma_0^2 \delta h (\xi)
Y_{lm} (\theta, \phi) X(t), \label{eq:deltanY}
\end{equation}
where the operator $\tilde{{\bf \nabla}}_T \equiv \hat{\theta}
(\partial /\partial \theta) + \hat{\phi} (1 / \sin \theta)
(\partial/\partial \phi)$.
Note that the variables $\xi$ and $t$ are separated
in above definitions of the perturbations, and so we consider only
``global'' perturbations \cite{Sari2000,Cox80}.  The function
$X(t)$ measures the amplitude of the perturbation relative to the
unperturbed values.

Substituting equations (\ref{eq:p_def_unpert})--(\ref{eq:n_def_unpert})
and (\ref{eq:deltagammaY})--(\ref{eq:deltanY}) into equations
(\ref{eq:deltamass})--(\ref{eq:deltaenergy}), we obtain
\begin{eqnarray}
& & \left[ q+2-k-m_1+\frac{2(m_1+2)}{(y-y_2)} \right] \frac{\delta
h} {h} - \frac{(y-2)}{2} \left( \frac{1}{g} \frac{1}{h} \frac{d
\delta h}{d\xi} \right) \nonumber \\ & & + \left[ \frac{-4(m_1+2)}
{(y-y_2)} + \frac{2(m_1+k-2)(y-d)} {(y-y_2)(y-2)} \right]
\frac{\delta g}{g} \nonumber \\ & & - \left( \frac{1}{g}
\frac{1}{g} \frac{d \delta g} {d\xi} \right) +
\frac{l(l+1)}{\Gamma_0^2} \delta v_T =0, \label{eq:perteqn1y}
\end{eqnarray}
\begin{eqnarray}
& & \left[-k-3m_1+q +\frac{2(m_1+2)(y-2)}{(y-y_2)} \right]
\frac{\delta f} {f} \nonumber \\ & & - \frac{(y+2)}{2} \left(
\frac{1}{g} \frac{1}{f} \frac{d \delta f}{d\xi} \right) \nonumber
\\ & & + \left[ 2q -\frac{2(m_1+2)(y-4)} {(y-y_2)} -
\frac{2(m_1+k-4)} {(y-y_2)} \right] \frac{\delta g}{g} \nonumber
\\ & & - (y-2) \left( \frac{1}{g} \frac{1}{g} \frac{d \delta g}
{d\xi} \right) =0, \label{eq:perteqn2y}
\end{eqnarray}
\begin{eqnarray}
& & \left[2(m_1+q+1) -\frac{2(m_1+k-4)}{(y-y_2)} \right] g\delta
v_T \nonumber \\ & & -(y-2) \frac{d \delta v_T} {d\xi}
-\frac{\delta f}{f} =0, \label{eq:perteqn3y}
\end{eqnarray}
\begin{eqnarray}
& & \left[\frac{2}{3}q -\frac{2}{3} \frac{(m_1+2)(y-4)}{(y-y_2)}
+\frac{2(m_1+k-4)}{(y-y_2)} \right. \nonumber \\ & & \left.
-\frac{8}{3} \frac{(m_1+k-2)(y-d)}{(y-y_2)(y-2)}\right]
\frac{\delta g} {g} -\frac{(y-2)}{3} \left( \frac{1}{g}
\frac{1}{g} \frac{d \delta g}{d\xi} \right) \nonumber \\ & &
+\left[ q -\frac{(m_1+k-4)(y-2)}{(y-y_2)} \right] \frac{\delta f}
{f} - \frac{(y-2)}{2} \left( \frac{1}{g} \frac{1}{f} \frac{d
\delta f}{d\xi} \right) \nonumber \\ & & -\frac{4}{3} \left[ q
-\frac{(m_1+k-2)(y-d)}{(y-y_2)} \right] \frac{\delta h} {h}
\nonumber \\ & & +\frac{2}{3} (y-2) \left( \frac{1}{g} \frac{1}{h}
\frac{d \delta h}{d\xi} \right) =0, \label{eq:perteqn4y}
\end{eqnarray}
where we have used equations
(\ref{eq:dfdxim1})--(\ref{eq:dhdxim1}) and the following relations
\begin{equation}
\frac{d \Gamma_0^2} {dt} = -m_1 \frac{ \Gamma_0^2} {t},
\label{eq:derGammat}
\end{equation}
\begin{equation}
\frac{\partial \xi} {\partial r} = - \frac{1}{t} \left[ \Gamma_0^2
+ \frac{1} {2(m_1+1)} \right], \label{eq:derxit}
\end{equation}
\begin{equation}
\frac{\partial \xi} {\partial t} = \frac{1}{t} \left[ \Gamma_0^2 -
\frac{m_1} {2(m_1+1)} -\xi (m_1+1) \right], \label{eq:derxir}
\end{equation}
\begin{equation}
\frac{d \rho_1} {dt} = -k \frac{\rho_1}{t} \left[1 -\frac{m_1}
{2(m_1+1)\Gamma_0^2} \right], \label{eq:derrhot}
\end{equation}
\begin{equation}
\frac{d n_1} {dt} = -k \frac{n_1}{t} \left[1 -\frac{m_1}
{2(m_1+1)\Gamma_0^2} \right], \label{eq:dernt}
\end{equation}
\begin{equation}
r=t \left[ 1 - \frac{1} {2(m_1+1) \Gamma_0^2} - \frac{\xi}
{\Gamma_0^2} \right]. \label{eq:rvst}
\end{equation}

Note that in deriving equations
(\ref{eq:perteqn1y})--(\ref{eq:perteqn4y}) we have assumed that there
is no perturbation in the external medium. Moreover, in order to
separate variables, $X(t)$ has to be a power law in time, $X(t)
\propto t^q$, where $q$ defines the temporal evolution of the
perturbation amplitude.  If the real part of $q$ is positive then the
perturbation grows, while if the real part of $q$ is
negative then the perturbation decays.

In equation (\ref{eq:perteqn1y}), the term $l(l+1)/\Gamma_0^2$ is
associated with causality, namely the fact that a perturbation can
only propagate at a speed $\lesssim c/\Gamma_0$ in the transverse
direction and hence expand across a maximum opening angle of $\sim
1/\Gamma_0$. Since $\Gamma_0$ is a function of time, it is not
possible to achieve a complete separation of variables for this
equation in contrast with the non-relativistic case.
However, for any constant value of $\Gamma_0$ we can still
calculate the power-law index for the growth of the perturbation,
$q$. These results are meaningful if we find $q>|m_1|$, so that
perturbations grow on a time scale shorter than the time scale for
changes in $\Gamma_0$. Therefore, the assumptions of variable
separation and fixed $\Gamma_0$ limit the generality of the
results. However, even if we find $q<|m_1|$, we should still be
able to gain an insight into some qualitative properties of the
perturbation amplitude evolution.

Equations (\ref{eq:perteqn1y})--(\ref{eq:perteqn4y}) are a complete
set of first-order differential equations for $\delta f$, $\delta g$,
$\delta h$, $\delta v_T$. After some algebraic manipulations, one may
write the equations for the first order terms $d \delta f /d\xi$, $d
\delta g /d\xi$, $d \delta h /d\xi$ and $d \delta v_T /d\xi$ in the
following matrix form
\begin{eqnarray}
& & (y^2-8y+4) \frac{d}{d\xi} \left( \begin{array}{c} \delta f \\
\delta g \\ \delta h \\ \delta v_T \end{array} \right) \nonumber
\\ & & = {\bf A}(q, k, l(l+1)/\Gamma_0^2, \xi)\left( \begin{array}{c}
\delta f \\ \delta g \\ \delta h \\ \delta v_T \end{array}
\right), \label{eq:perteqnmatrix}
\end{eqnarray}
where ${\bf A}$ is a $4\times 4$ matrix. Note that
$(y^2-8y+4)=(y-y_1)(y-y_2)$. Thus, the solutions for the perturbation
variables must pass the same singular point (or the sonic line),
$y_1=4-2\sqrt{3}$, as the unperturbed variables.  Therefore,
the value of $q$ can be found by requiring that the solutions pass through
the singular point $y_1$. This is very similar to the non-relativistic
case \cite{Sari2000}.

In order to numerically integrate the differential equations
(\ref{eq:perteqnmatrix}) and derive $q$ we need to specify the
boundary conditions at the shock front when the shock is perturbed.
Since the relativistic jump conditions across the shock front must be
satisfied, we have
\begin{equation}
p=\frac{2}{3}\Gamma^2 \rho_1 , \label{eq:jump1}
\end{equation}
\begin{equation}
n'=2\Gamma^2 n_1 , \label{eq:jump2}
\end{equation}
\begin{equation}
\gamma^2=\frac{1}{2}\Gamma^2 , \label{eq:jump3}
\end{equation}
where $\Gamma$ is the Lorentz factor of the perturbed shock front.
By linearizing these boundary conditions with respect to the
perturbed quantities, we find
\begin{equation}
\delta p + \left( \frac{\partial p_0} {\partial r} \right) \delta
R = \frac{4}{3} \Gamma_0^4 \rho_1 \frac{d}{dt} \delta R -
\frac{2k}{3} \Gamma_0^2 \rho_1 \frac{\delta R}{R_0},
\label{eq:bound1pert}
\end{equation}
\begin{equation}
\delta n' + \left( \frac{\partial n'_0} {\partial r} \right)
\delta R = 4 \Gamma_0^4 n_1 \frac{d}{dt} \delta R - 2k \Gamma_0^2
n_1 \frac{\delta R}{R_0}, \label{eq:bound2pert}
\end{equation}
\begin{equation}
\delta \gamma^2 + \left( \frac{\partial \gamma_0^2} {\partial r}
\right) \delta R = \Gamma_0^4 \frac{d}{dt} \delta R ,
\label{eq:bound3pert}
\end{equation}
where $\delta R$ is the deviation of the perturbed shock radius
$R$ from the unperturbed shock radius $R_0$. In deriving equations
(\ref{eq:bound1pert})--(\ref{eq:bound3pert}), we used the
relations $\rho_1 \propto R^{-k}$ and
\begin{equation}
\delta \Gamma^2 = 2 \Gamma_0^4 \frac{d}{dt} \delta R ,
\label{eq:deltaGamma}
\end{equation}
where $\delta \Gamma^2$ is the deviation of the square of the
perturbed shock Lorentz factor $\Gamma^2$ from the square of the
unperturbed shock Lorentz factor $\Gamma_0^2$.

We now define
\begin{equation}
\delta R (\theta, \phi, t) = \eta \frac{1}{\Gamma_0^2} R_0 Y_{lm}
(\theta, \phi) X(t), \label{eq:deltaR}
\end{equation}
where $\eta$ is a scale factor that can have an arbitrary value;
for convenience we set $\eta =1$. Substituting equations
(\ref{eq:deltaR}),
(\ref{eq:p_def_unpert})--(\ref{eq:n_def_unpert}),
(\ref{eq:deltagammaY}), (\ref{eq:deltapY}) and (\ref{eq:deltanY})
into equations (\ref{eq:bound1pert})--(\ref{eq:bound3pert}), we
find
\begin{equation}
\delta f = \frac{df} {d \xi} + 2(m+q+1), \label{eq:pertbound1}
\end{equation}
\begin{equation}
\delta h = \frac{dh} {d \xi} + 2(m+q+1), \label{eq:pertbound2}
\end{equation}
\begin{equation}
\delta g = \frac{dg} {d \xi} + 2(m+q+1). \label{eq:pertbound3}
\end{equation}

Another boundary condition results from the requirement that the
tangential velocities must be continuous across the shock front,
yielding
\begin{equation}
\delta {\bf v}_T = -\frac{1}{R_0} \tilde{ {\bf \nabla}}_T \delta R
. \label{eq:jumpvt}
\end{equation}
Substituting equations (\ref{eq:deltavtY}) and (\ref{eq:deltaR}) into
this equation, we get
\begin{equation}
\delta v_T =1 . \label{eq:pertbound4}
\end{equation}
Equations (\ref{eq:pertbound1})--(\ref{eq:pertbound3}) and
(\ref{eq:pertbound4}) are the four boundary conditions necessary
to solve the perturbation equations.

\subsection{Numerical results}

Based on the derivations presented in the previous subsection, we
may now examine the stability of different modes for different
values of $k$. As a particular example, we consider the case of
$k=5.5$. We derive $q$ for different values of the mode wavenumber
$l$ by integrating the differential equations
(\ref{eq:perteqnmatrix}) from the shock front to its interior and
requiring that the solutions pass through the singular point. The
results are shown in Figure 2. The top panel shows the real
component of $q$ (${\rm Re}[q]$) as a function of
$\sqrt{l(l+1)}/\Gamma_0$, where $\Gamma_0$ is treated as a scaling
factor. As mentioned before, ${\rm Re}[q]$ determines the growth
rate of the perturbation. In the bottom panel, we plot the
imaginary component of $q$ (${\rm Im}[q]$), which provides the
oscillation frequency of the perturbation.
Figure 2 separates the
behavior of $q$ into three different regimes:
\begin{itemize}

\item{} In the regime of
small $l/\Gamma_0$ ($0< l/\Gamma_0 < 0.87$), $q$ is a real number
and ${\rm Re}[q]$ is positive, implying that the perturbation grows
monotonically in time. The value of ${\rm Re}[q]$ increases as $l$
increases. Note that $q$ vanishes in the limit of $l\rightarrow 0$. This
result can be derived analytically by comparing two unperturbed
spherical solutions with different parameters.

\item{}In the regime of
intermediate $l/\Gamma_0$ ($0.87 < l/\Gamma_0 < 17$), $q$ is a
complex number and ${\rm Re}[q]$ is positive, implying that  the
perturbation grows while oscillating. As $l$ increases the real
part ${\rm Re}[q]$ decreases while the imaginary part ${\rm
Im}[q]$ increases. Note that the transition between the real and
imaginary solutions for $q$ occurs at $l/\Gamma_0 \sim 0.87$. This
result follows from causality. When the wavelength of the
perturbation ($\sim 1/l$) is smaller than $1/\Gamma_0$, the
maximum angular separation of two regions that can interact with
each other, the perturbation can oscillate.

\item{} Finally in the regime
of large $l/\Gamma_0$ ($l/\Gamma_0 > 17$), $q$ is a complex number
and ${\rm Re}[q]$ is negative, implying that the perturbation decays
while oscillating. The value of ${\rm Re}[q]$ decreases (so the
absolute value of ${\rm Re}[q]$ increases) as $l$ increases while
the value of ${\rm Im}[q]$ increases as $l$ increases.
\end{itemize}

\begin{figure}[!t]
\centerline{\psfig{figure=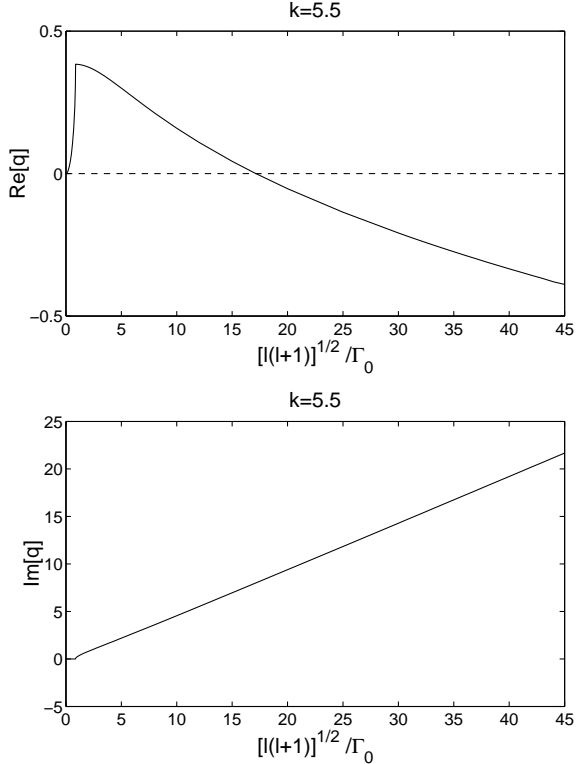,width=3.0in}} \caption[]{
Perturbation growth rate, $q$, as a function of $\sqrt{l(l+1)} /
\Gamma_0$. The upper and lower panels show ${\rm Re}[q]$ and ${\rm
Im}[q]$ respectively. } \label{figure2}
\end{figure}

The actual evolution of a perturbation is shaped by the fact that
$\Gamma_0$ increases with time as $\Gamma_0^2 \propto t^{-m_1}$.  If
initially the wavenumber of the perturbation is sufficiently large so
that it is in the regime of large $l/\Gamma_0$, the perturbation will
start to decay while oscillating. As time progresses, $l/\Gamma_0$
decreases and so both $\mid {\rm Re}[q]\mid$ and ${\rm Im}[q]$
decrease, the perturbation decays with slower speed and oscillates on
longer timescales. As soon as the perturbation enters the regime of
intermediate $l/\Gamma_0$, it starts to grow slowly over time and
oscillate on even longer timescales. The growth rate slowly increases
over time, but is always limited by the rather small upper bound,
${\rm Re}[q] \lesssim 0.38$. Eventually, the perturbation enters the
regime of small $l/\Gamma_0$ and grows slowly without oscillating. As
$t$ increases, the growth rate approaches zero, and so the
perturbation saturates.  Therefore, perturbations with large
wavenumbers (short wavelengthes) grow when $1\lesssim l/\Gamma_0\lesssim
10$ only by a modest factor.  In the case of intermediate wavenumbers,
the perturbation goes through the two regimes of intermediate and
small $l/\Gamma_0$.  Therefore it grows slowly with some initial
oscillations, but soon afterwards it stops oscillating and
saturates. Perturbations with small wavenumbers stay in the regime of
small $l/\Gamma_0$. The perturbation grows slowly without oscillating
at the beginning but soon saturates.

The above results remain qualitatively the same for all values of
$k>4.134$.

\section{Numerical Simulations}

We have verified the above behavior by a direct integration of the
partial different equations which determine the evolution of the
perturbation variables, without  assuming separability of the
solutions with respect to $\xi$ and $t$. Instead of equations
(\ref{eq:deltagammaY}) -- (\ref{eq:deltanY}), we redefined the
perturbation variables as
\begin{equation}
\delta \gamma^2 (r, \theta, \phi, t) = \frac{1}{2} \Gamma_0^2
\delta g (\xi, t) Y_{lm} (\theta, \phi), \label{eq:deltagammaY_2}
\end{equation}
\begin{equation}
\delta {\bf v}_T (r, \theta, \phi, t) = -\frac{1}{\Gamma_0^2}
\delta v_T (\xi, t) \tilde{ {\bf \nabla}}_T Y_{lm} (\theta, \phi),
\label{eq:deltavtY_2}
\end{equation}
\begin{equation}
\delta p (r, \theta, \phi, t) = \frac{2}{3} \rho_1 \Gamma_0^2
\delta f (\xi, t) Y_{lm} (\theta, \phi), \label{eq:deltapY_2}
\end{equation}
\begin{equation}
\delta n' (r, \theta, \phi, t) = 2 n_1 \Gamma_0^2 \delta h (\xi,
t) Y_{lm} (\theta, \phi). \label{eq:deltanY_2}
\end{equation}
Equations (\ref{eq:perteqn1y}) -- (\ref{eq:perteqn4y}) were then
replaced by four partial differential equations (PDEs) for the
perturbation variables $\delta f (\xi, t)$, $\delta g (\xi, t)$
$\delta h (\xi, t)$ and $\delta v_T (\xi, t)$. We then solved for
the evolution of these perturbation variables by numerically
integrating the PDEs with appropriate initial values.
In our numerical simulations, the outer boundary ($\xi=0$) is the
shock front where the shock jump conditions are assumed to be
satisfied. We can still define $\delta R$ as in equation
(\ref{eq:deltaR}). Then at the outer boundary the perturbation
variables satisfy
\begin{equation}
\delta f (0,t)= \frac{df} {d \xi} X(t) + 2(m_1 +1)X(t) +
2t\frac{dX(t)}{dt}, \label{eq:pertbound1_2}
\end{equation}
\begin{equation}
\delta h (0,t)= \frac{dh} {d \xi} X(t) + 2(m_1 +1)X(t) +
2t\frac{dX(t)}{dt}, \label{eq:pertbound2_2}
\end{equation}
\begin{equation}
\delta g (0,t)= \frac{dg} {d \xi} X(t) + 2(m_1 +1)X(t) +
2t\frac{dX(t)}{dt}, \label{eq:pertbound3_2}
\end{equation}
\begin{equation}
\delta v_T (0,t)= X(t) . \label{eq:pertbound4_2}
\end{equation}
The inner boundary is chosen to be sufficiently large so as to
cover the entire similarity region which is bounded by a inner
$C_+$ characteristic. This way, the values of the perturbation
variables at the inner boundary can not affect the shock front.

The numerical simulations gave us the same behavior for the
perturbations as the previously mentioned analytical results for the
growth rate $q$. In Figure 3, we show the evolution of $X(t)$ in a
numerical simulation with $k=5.5$ and $\l/\Gamma_0=75$. We plot $X(t)$
in the range $t \in [1, 10]$ in the top panel and $t \in [1, 100]$ in
the bottom panel. Note that $X(t)$ describes the relative displacement
of the perturbed shock radius from the unperturbed value and we set
its initial value to be $X(t=1)=1.0$.  From Figure 3 we see that $X$
oscillates over time over increasingly longer timescales and stops
oscillating at late times. Its amplitude first decreases, then slowly
increases and finally saturates.  Overall it grows by a factor of
$\sim 10$.  These results are consistent with the previous discussion
on the three regimes for the evolution of the perturbations.

\begin{figure}[!t]
\centerline{\psfig{figure=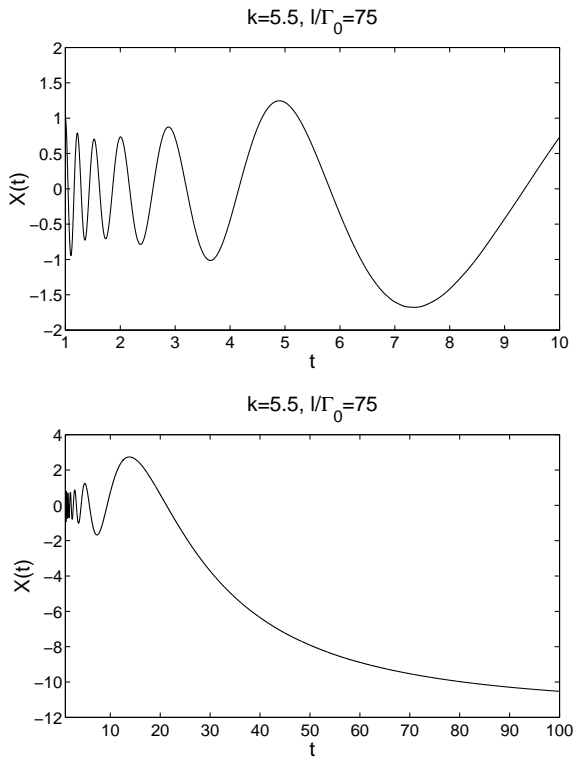,width=3.0in}} \caption[]{
Evolution of $X(t)$ for $k=5.5$ and $\l/\Gamma_0=75$. The upper
and lower panels show $t \in [1, 10]$ and $t \in [1, 100]$
respectively. } \label{figure3}
\end{figure}

\section{Conclusions}

We have derived the self-similar solutions for an ultra-relativistic blast
wave in an external medium with a density profile $\rho_1 \propto r^{-k}$ and
$k>4$. The solutions exist for $k$ larger than a critical value
$k_c=4.134$. They describe the flow in the self-similar region bounded by
the shock front and a $C_+$ characteristic. The shock front accelerates
with Lorentz factor $\Gamma^2 \propto t^{-m_1}$ and $m_1<-1.134$, while the
$C_+$ characteristic accelerates with Lorentz factor $\gamma^2 \propto
t^{2}$. The energy and mass contained inside the self-similar region
approach constant values as time diverges.


We have found that at large wavenumbers the perturbations first decay,
then grow slowly over time and eventually saturate. The initial decay
and the intermediate growth are accompanied by temporal
oscillations. These small wavelength perturbations grow when $1\lesssim
l/\Gamma_0\lesssim 10$ with an overall factor of $\sim 10$.  At
intermediate wavenumbers, the perturbations first grow slowly and then
saturate. The initial growth is also accompanied by temporal
oscillations. At small wavenumbers the perturbations grow
monotonically in time but soon saturate. Our results also apply to
expanding relativistic jets as long as the opening angle of the jet is
larger than the inverse of its Lorentz factor.

In the collapsar model of gamma-ray bursts, a collimated relativistic
outflow is generated due to the collapse of the core of a massive
star. The outflow approaches the stellar envelope at a modest
semi-relativistic speed but is expected to accelerate significantly
across the sharp density gradient at the surface of the
star\cite{Zhang2002}.  Our results indicate that in the breakout phase
perturbations are close to being stable in spherical symmetry.  It is
still possible, however, that the lateral expansion of the jet at
breakout would be accompanied by instabilities. These instabilities
may produce variations in the Lorentz factor of the jet needed in the
internal shock model. They may also be responsible for the complex
light curves observed in most GRBs.  Current numerical
simulations\cite{Zhang2002} lack adequate resolution at the stellar
surface to follow the shock breakout and confirm the instabilities. We
leave a detailed study of the instabilities associated with the
lateral expansion of the jet for future work.

\paragraph*{Acknowledgments.}
This work was supported in part by grants from the Israel-US BSF
(BSF-9800343) and NSF (AST-0071019, AST-0204514).

\end{multicols}
\end{document}